\documentstyle[a4,epsf,amstex]{article}
 
\newcommand{\bee}{\begin{equation}}
\newcommand{\ee}{\end{equation}}
\newcommand{\beea}{\begin{eqnarray}}
\newcommand{\eea}{\end{eqnarray}}

\begin{document}
\thispagestyle{empty}
\parskip=12pt
\raggedbottom
 
\def\mytoday#1{{ } \ifcase\month \or
 January\or February\or March\or April\or May\or June\or
 July\or August\or September\or October\or November\or December\fi
 \space \number\year}
\noindent
\hspace*{9cm} BUTP--98/4\\
\vspace*{1cm}
\begin{center}
{\LARGE Lattice QCD without tuning, mixing and current renormalization }
\footnote{Work supported in part by Schweizerischer Nationalfonds.}
 
\vspace{1cm}
Peter Hasenfratz 
\\
Institute for Theoretical Physics \\
University of Bern \\
Sidlerstrasse 5, CH-3012 Bern, Switzerland

\vspace{0.5cm}

\vspace{0.5cm}
\mytoday \time \\ \vspace*{0.5cm}

\nopagebreak[4]
   
\begin{abstract}
The classically perfect action of QCD requires no tuning to
get the pion massless in the broken phase: the critical bare mass $m_q^c$ is
zero. Neither the vector nor the flavour non-singlet axial vector currents
need renormalization. Further, there is no mixing between four-fermion
operators in different chiral
representations. The order parameter of chiral symmetry requires, however a
subtraction which is given here explicitly. These results are based on the
fact that the fixed point action satisfies the Ginsparg-Wilson remnant chiral
symmetry condition. On chiral symmetry related questions any other local
solution of this condition will produce similar results.
\end{abstract}
 
\end{center}
\eject

\section{Introduction}
Lattice regularized, local QCD actions break chiral symmetry under 
general conditions \cite{NN}. Wilson fermions \cite{KW} have 
a dimension five
symmetry breaking operator whose effect on the physical predictions goes to
zero in the continuum limit. It leaves, however its trace behind in the form
of additive quark mass renormalization, axial current renormalization and
mixing between operators in nominally different chiral representations. There
is a significant recent progress in calculating these renormalizations in a
theoretically controlled non-perturbative way \cite{NONP}. 
Nevertheless, the situation is
not really pleasing theoretically and the technical difficulties are
also significant.

Staggered fermions \cite{KSU} keep a part of chiral symmetry intact which
offers another possibility to study problems where chiral symmetry is
essential \cite{AOSH}. On the other hand, staggered fermions do not solve the
doubling problem and constructing operators with correct quantum numbers is
far from trivial. 

A rather different method to overcome the problems with chiral symmetry is the
domain wall fermion \cite{KAP,SONI}, or the overlap formalism
\cite{OVER}. Recent results on kaon matrixelements are nice and promising
\cite{SONI}. Similarly to the staggered fermions, the domain wall fermions
were motivated by a single issue: to solve the problems of chiral symmetry in
the fermion sector.

This paper is part of a project to construct a lattice
formulation for QCD which performs well in every respect
both in the gauge and in the fermion sectors, including
classical solutions, topology, cut-off effects and chiral symmetry.
The fixed point (FP) action ($m_q=0$) and the actions on the trajectory along
the {\em mass} direction ($m_q \not=0$), which are local and
determined by saddle point equations, are classically 
perfect \cite{HN,BB,UW,PH}. Among others,
the FP action is perfect concerning the classical solutions leading to scale
invariant instantons \cite{HN,BB,BBHN,ROTOR,FN}
and fermionic chiral zero modes satisfying the index theorem
\cite{ZERO}. Even more, as
we discuss here, these actions are {\em quantum perfect} what concerns chiral
symmetry \cite{4}. The problem of constructing the FP action and the mass
trajectory by solving the corresponding classical saddle-point equations
requires skill and patient. The real difficulty is, however to find a 
parametrization for these actions which is sufficiently precise and, at the
same time, does not make the simulations too expensive. Examples in $d=2$ 
show \cite{HN,BBHN,SPI} that with a limited number of
couplings a parametrization can be achieved which performs excellently in
simulations and other numerical checks. In d=4, although the preliminary
results look promising,
the parametrizations studied
until now are admittedly rather primitive \cite{BB,PA}.

As the discussion above shows, there exist several possibilities to cope with
the problems of chiral symmery. It might turn out on the long run that none of
the new theoretical ideas can compete with the direct approach of
using Wilson fermions and calculating renormalization factors, 
mixing coefficients in different processes and driving the system close to the
continuum limit to kill the cut-off effects and to avoid other 
problems \cite{AOKI}. The author hopes that the solution will be more
appealing. 
 
It is somewhat surprising that the FP action respects chiral symmetry in its
predictions, since -- 
complying with the Nielsen-Ninomiya theorem \cite{NN} --, it breaks 
chiral symmetry
explicitly. The basic observation is that this breaking is realized in a very
special way. The FP action satisfies the equation \cite{PH,ZERO}
\footnote{We denote the commutator and anticommutator by 
[\,,\,]
  and \{ \,,\, \}, respectively.}
\begin{equation}
\label{1}
\frac {1}{2}\{ h^{-1}_{n n'} (U), \gamma^5 \} \gamma^5 = 
R_{n n'}(U) \,,
\end{equation}
or equivalently
\begin{equation}
\label{2}
\frac {1}{2}\{ h_{n n'}, \gamma^5 \} = 
(h \gamma^5 R h)_{n n'} \,.
\end{equation} 
Here $h(U)^{-1}$ is the FP quark propagator over the background gauge field $U$
and the matrix $R_{nn'}$ is trivial in Dirac space. Both $h(U)_{nn'}$ and
$R(U)_{nn'}$ are local. The precise form of $R$ depends on the block
transformation whose FP we are considering.

It has been observed a long time ago that eq.~(\ref{2}) with a local $R$ is
the mildest way a local lattice action can break chiral symmetry
\cite{GW}. Along the mass trajectory ($m_q \not= 0$) eq.~(\ref{2}) is modified
as 
\begin{equation}
\label{3}
\frac {1}{2}\{ {\hat h}_{n n'}, \gamma^5 \} = 
({\hat h} \gamma^5 R {\hat h})_{n n'}+
m_q \Sigma^5_{n n'} \,,
\end{equation} 
where we denoted the action along the mass trajectory by 
$\hat{h}$, $\hat{h} \rightarrow h$ as $m_q \rightarrow 0$. 
The operator $\Sigma^5$ is a local
pseudoscalar density which goes to $\gamma^5 \delta_{nn'}$ in the formal
continuum limit. The results of this paper are based on eq.~(\ref{3}), hence
any local solution of this equation will lead to the same results. It is
interesting in this context that in a recent paper \cite{NE} another solution
(growing out of the overlap formalism)
of eq.~(\ref{1}), which seems to be unrelated to renormalization group
considerations, was presented. This is an interesting development, if it
can be shown that the Dirac operator in \cite{NE}
is local over any gauge
configuration.
 
Since the locality of $h, R$ and the currents is basically important
in deriving the following results, let us discuss briefly what is meant by
this notion. 
An operator (like action density,
topological charge density, current, etc.) on the lattice is local if it has
an extension of $O(a)$: the coupling between the fields in the operator at a
distance $n$ decays exponentially as $\exp(-\gamma n)$, with $\gamma=O(1)$, or
it is identically zero beyond a certain $O(a)$ range. In the continuum limit
($a \to 0$) the extension measured in physical units goes to zero. A quantum
field theory with non-local interactions looses renormalizability and
universality in general -- this would be certainly too high a price for an
'improvement'.

The remark has been made already in \cite{GW} that the soft-pion theorems are
expected to remain valid if eq.~(\ref{2}) is satisfied. The intuitive reason
is that the two $h$ factors on the r.h.s. of eq.~(\ref{2}) will cancel the two
propagators which connect this term to other operators in the matrixelement
producing only a contact term in Ward identities. On the other hand, in current
algebra relations, in Ward identities related to mixing and elsewhere, contact
terms are relevant, so it is necessary to study in detail what eq.~(\ref{3})
implies on chiral symmetry. Many of our considerations rely on the seminal
paper of Bochicchio et al. \cite{CCH} where the chiral symmetry properties of
Wilson fermions are discussed.

\section{The vector and axial currents}

The classically perfect fermion action has the form
\begin{equation}
\label{4}
S_{\rm f}^{\rm FP}(\bar{\psi},\psi,U)=
\sum_{m',n'}\bar{\psi}_{m'} {\hat h}_{m' n'}(U)\psi_{n'}\,.
\end{equation} 
The Dirac, flavour and colour indices are suppressed in eq.~(\ref{4}). The
Dirac operator $\hat h$ is assumed to be flavour independent. The flavour
generators are denoted by $\tau^a, a=1,\dots,N_f^2-1$, 
$tr(\tau^a \tau^b)=\frac{1}{2} \delta_{ab}$. 

As usual, the axial currents are constructed from the chiral symmetric part of
the action using 
${\hat h}_{\rm SYM}=\frac{1}{2} [{\hat h},\gamma^5] \gamma^5$. 
There is a certain freedom in constructing the
vector currents. Using the full action, conserved vector currents are
obtained. On the other hand, the asymmetry between the vector and axial case
generates strange terms in current algebra relations and it requires extra
work to show that they go away. We decided to follow \cite{CCH} and construct
both currents with ${\hat h}_{\rm SYM}$. We get \footnote{We use the 
standard sign
convention for the currents as opposed to that in \cite{GW,ZERO}. We use the
notation ${\bar \nabla}_\mu f(n)=f(n)-f(n-\hat\mu)$ .} 
\begin{equation}
\label{5}
\bar{\nabla}_\mu V_\mu^a (n)= 
\bar{\psi}_n \tau^a \left( {\hat h}_{\rm SYM} \psi \right)_n - 
\left( \bar{\psi} {\hat h}_{\rm SYM} \right)_n \tau^a \psi_n \,, 
\end{equation} 
\begin{equation}
\label{6}
\bar{\nabla}_\mu A_\mu^a (n)= 
- \bar{\psi}_n \tau^a \left( \gamma^5 {\hat h}_{\rm SYM} \psi \right)_n - 
\left( \bar{\psi} {\hat h}_{\rm SYM}\gamma^5 \right)_n \tau^a \psi_n \,. 
\end{equation}
None of these currents are conserved, at least not without further
considerations, 
if the equations of motion
${\hat h} \psi =0$ is used.

In order to find the currents themselves we introduce the flavour gauge
matrices $W_\mu(n)= 1 + i w_\mu^a(n)\tau^a +\dots$ and extend the
product of colour $U$ matrices in eq.~(\ref{4}) along the paths between the
fermion offsets $m',n'$ by the product of $W$ matrices along the same
paths. Then the vector current is defined by
\begin{equation}
\label{7}
V_\mu^a(n) = -i
\sum_{m',n'} \bar{\psi}_{m'} 
\left. \frac{\delta}{\delta w_\mu^a(n)} 
(\hat{h}_{\rm SYM})_{m' n'}(U,W) \psi_{n'} \right|_{w=0} \,.
\end{equation} 
This current satisfies eq.~(\ref{5}) \cite{PHF}. Every path in
$\left( {\hat h}_{\rm SYM} \right)_{m'n'}$ which goes through the link
$(n,n+\hat{\mu})$ gives a contribution to $V_\mu^a(n)$ which is equal to the
contribution to $\left( {\hat h}_{\rm SYM} \right)_{m'n'}$ 
times $(s_+ - s_-)\tau^a$,
where $s_+ (s_-)$ is the number this path runs through $(n,n+\hat{\mu})$ in
positive (negative) direction. Eq.~(\ref{7}) can also be written as 
\begin{equation}
\label{8}
V_\mu^a(n) =
\sum_{m' n'} \bar{\psi}_{m'} \tau^a \Gamma_\mu(m',n';n;U)
\psi_{n'} \, ,
\end{equation} 
where $\Gamma_\mu$ is trivial in flavour space. Similar considerations lead to
the axial current 
\begin{equation}
\label{9}
A_\mu^a(n) =
\sum_{m' n'} \bar{\psi}_{m'} \tau^a \Gamma_\mu^5(m',n';n;U)
\psi_{n'} \, ,
\end{equation} 
where 
\begin{equation}
\label{10}
\Gamma_\mu^5 = \Gamma_\mu \gamma^5 = - \gamma^5 \Gamma_\mu \, .
\end{equation}

\section{The basic Ward identity}

We shall repeatedly use the following Ward identity which can be obtained by
changing integration variables in the QCD path integral (see, for example
\cite{CCH}) 
\begin{eqnarray}
\label{11}
i \langle \frac{\delta}{\delta \epsilon_n^a} {\cal O}(x_1,\dots,x_j)\rangle  &=&
\langle {\cal O}(x_1,\dots,x_j)\bar{\nabla}_\mu A_\mu^a (n)\rangle  -\nonumber
\\
& & \langle {\cal O}(x_1,\dots,x_j) X^a(n)\rangle -  \\
& &  \langle {\cal O}(x_1,\dots,x_j)2 m_q P^a(n)\rangle \nonumber  \,,
\end{eqnarray}  
where ${\cal O}$ is some product of local operators, $\delta {\cal O}$ is its
change under a local chiral transformation with the infinitesimal parameter
$\epsilon_n^a$, $P^a(n)$ is a pseudoscalar density
\begin{equation}
\label{12}
P^a (n)= \frac{1}{2}[
\bar{\psi}_n \tau^a \left( \Sigma^5 \psi \right)_n + 
\left( \bar{\psi} \Sigma^5 \right)_n \tau^a \psi_n ] \,, 
\end{equation}  
while $X^a$ comes from the remnant chiral symmetry condition eq.~(\ref{2}) 
\begin{equation}
\label{13}
X^a (n)= 
\bar{\psi}_n \tau^a ({\hat h}\gamma^5 R {\hat h} \psi )_n + 
( \bar{\psi}{\hat h}\gamma^5 R {\hat h} )_n \tau^a \psi_n \,. 
\end{equation}
In deriving eq.~(\ref{11}) we used eq.~(\ref{6}), 
$\hat {h}_{\rm SYM}=\hat {h}-\frac{1}{2}\{ \hat{h},\gamma^5 \} \gamma^5 $ 
and eq.~(\ref{3}). For
later use we introduce the flavour singlet scalar density $S(n)$
\begin{equation}
\label{14}
S(n)= 
\frac{1}{2}[ \bar{\psi}_n  \left( \Sigma \psi \right)_n + 
\left( \bar{\psi} \Sigma \right)_n  \psi_n ] 
\end{equation}
with
\begin{equation}
\label{15}
\Sigma = \frac{1}{2}\{\Sigma^5,\gamma^5 \}\,.
\end{equation}

\section{The limit $m_q \rightarrow 0$ and the order parameter of chiral
symmetry} 

The following considerations are valid in the broken phase in general, not
only in the continuum limit. We shall consider the Ward identity
eq.~(\ref{11}) with ${\cal O}=P^b(x)$ and sum over $n$. After some algebra we
obtain
\begin{equation}
\label{16}
\sum_n i \langle \frac{\delta}{\delta \epsilon_n^a} P^b(x)\rangle  =
-\delta_{ab} \frac{1}{N_f} \langle S(x)\rangle  \,,
\end{equation}
\begin{equation}
\label{17}
-\sum_n \langle P^b(x) X^a(n)\rangle  = \delta_{ab} \langle {\rm tr}^{\rm DC}\frac{1}{2}
\{\Sigma^5(U),\gamma^5 R(U)\}_{xx}\rangle \,, 
\end{equation}
where the trace ${\rm tr}^{\rm DC}$ in eq.~(\ref{17}) is over Dirac and colour
space. We used the notations introduced in eqs.~(\ref{12}-\ref{15}). The last
term in eq.~(\ref{11}) is dominated by the pion state for small
$m_q$: 
\begin{equation}
\label{18}
 -2 m_q \langle P^b(x) \sum_n P^a(n)\rangle  = 
2 m_q \delta_{ab} \frac{|\alpha|^2}{m_\pi^2}\,, \qquad (m_q \rightarrow 0) \,,
\end{equation}
where we denoted the pion--vacuum matrixelement of the pseudoscalar density
$P$ by $\alpha$. We shall see later (eq.~(\ref{30}))
that $|\alpha|^2=f_\pi^2(m_\pi^2/m_q)^2/4$
in the continuum limit, but it is sufficient at this moment to observe that
$\alpha$ is finite as $m_q \rightarrow 0$. 
Eqs.~(\ref{11},\ref{16},\ref{17},\ref{18}) give
\begin{equation}
\label{19}
- \langle S(x) + N_f{\rm tr}^{\rm DC}\frac{1}{2}
\{\Sigma^5,\gamma^5 R\}_{xx}\rangle  =
2 N_f \frac{m_q}{m_\pi^2}|\alpha|^2 \,, \qquad (m_q \rightarrow 0)\,.
\end{equation}
We shall show now that the combination on the l.h.s. is one of the possible
order parameters of chiral symmetry: it is zero in perturbation theory, or
more generally in the limit $\lim_{V \to \infty} \lim_{m_q \to 0}$ (in this
order). If this order parameter picks up a non-zero expectation value in the
limit $\lim_{m_q \to 0} \lim_{V \to \infty}$ due to non-perturbative effects
then eq.~(\ref{19}) implies
\begin{equation}
\label{191}
m_\pi^2 \sim  m_q, \qquad (m_q \rightarrow 0)\,,
\end{equation}
i.e. the critical quark mass is zero. There is no tuning.

Let us first construct the simplest order parameter. Consider
\begin{equation}
\label{20}
\langle {\bar\psi}_x \psi_x\rangle _{\rm sub} =
\langle {\bar\psi}_x \psi_x + 4 N_f {\rm tr}^{\rm C} R(U)_{xx}\rangle 
\end{equation}
in the limit $m_q \rightarrow 0$ in a finite volume. Integrating out the
fermions we get
\begin{equation}
\label{21}
\langle {\bar\psi}_x \psi_x\rangle _{\rm sub} =
\langle {\rm tr}^{\rm DFC}(-h_{xx}^{-1} + R_{xx})\rangle  \,,
\end{equation} 
where the trace is over Dirac, flavour and colour space. The r.h.s. of
eq.~(\ref{21}) is zero as can be seen easily by taking the trace of
eq.~(\ref{1}). For specific block transformations ('blocking out of continuum'
\cite{BW}) $R_{xx}=1/\kappa$, where $\kappa$ is a parameter entering the
blocking procedure. In this case the order parameter is obtained from
$\langle {\bar \psi}_x \psi_x\rangle $ by simply adding the constant $4N_fN_c/\kappa$.

Eq.~(\ref{1}) allows to construct other order parameters also. Multiplying
eq.~(\ref{1}) by $\Sigma^5 \gamma^5$ from the left and adding to it the
product from the right, after taking the trace
the expectation value on the l.h.s. of eq.~(\ref{19}) is
obtained as another definition of the order parameter. This is what we wanted
to show.

\section{Current renormalization}

Current algebra relations which are non-linear in the currents, are very
convenient for the study the renormalization of the currents. Consider the Ward
identity, studied also by Bochicchio et al. \cite{CCH} in this context, in the
continuum
\begin{eqnarray}
\label{22}
\partial_\mu^x \langle A_\mu^a(x) A_\nu^b(y) V_\rho^c(z)\rangle  &=&
2m_q \langle  P^a(x) A_\nu^b(y) V_\rho^c(z)\rangle  + \nonumber \\
& & i f^{abd}\delta(x-y) \langle V_\nu^d(y)V_\rho^c(z)\rangle  + \\
& & i f^{acd}\delta(x-z) \langle A_\nu^b(y)A_\rho^d(z)\rangle  \,. \nonumber
\end{eqnarray}
Integrating over $x$, we obtain
\begin{eqnarray}
\label{23}
0 = \int dx \,2m_q \langle  P^a(x) A_\nu^b(y) V_\rho^c(z)\rangle  &+&
i f^{abd} \langle V_\nu^d(y)V_\rho^c(z)\rangle   \nonumber \\
&+ & i f^{acd} \langle A_\nu^b(y)A_\rho^d(z)\rangle \,. 
\end{eqnarray}
We shall consider eq.~(\ref{23}) for $|y-z|$ being a physical distance. With
this condition we assure that the
communication between the points $y$ and $z$ goes through physical intermediate
states.
We work out the Ward identity corresponding to eq.~(\ref{23}) on the
lattice. Consider the general Ward identity in eq.~(\ref{11}) with
\begin{equation}
\label{24}
{\cal O} \rightarrow A_\nu^b(y) V_\rho^c(z) \,,
\end{equation}
where $|y-z|$ is much larger than the lattice unit $a$. We get \footnote{The
simple relations in eqs.~(\ref{25},\ref{26})  would receive additional terms
had we used the conserved form of the vector current. Although they can be
shown not to influence the final conclusions, their presence would complicate
our considerations.}
\begin{equation}
\label{25}
\sum_n i  \frac{\delta}{\delta \epsilon_n^a} V_\rho^c(z) =
i f^{acd} A_\rho^d(z) \,,
\end{equation}
\begin{equation}
\label{26}
\sum_n i  \frac{\delta}{\delta \epsilon_n^a} A_\nu^b(y) =
i f^{abd} V_\nu^d(y) \,.
\end{equation}

Using eqs.~(\ref{24},\ref{25},\ref{26}) in eq.~(\ref{11}), the continuum Ward
identity eq.~(\ref{23}) is reproduced if
\begin{equation}
\label{27}
\langle \sum_n X^a(n) A_\nu^b(y) V_\rho^c(z)\rangle  = 0 \,.
\end{equation}
Eqs.~(\ref{8},\ref{9},\ref{13}) give after integrating out the fermions
\begin{equation}
\label{28}
-2 {\rm Tr}\left( (\tau^a\tau^b\tau^c)(\Gamma_\rho(z)\gamma^5 R 
\Gamma_\nu^5(y)+
\Gamma_\nu^5(y)\gamma^5 R \Gamma_\rho(z)){\hat h}^{-1}\right)\,,
\end{equation}
where Tr is a trace over all the indices, including space.
Since $R, \Gamma_\rho$ and
$\Gamma_\nu^5$ are local, eq.~(\ref{28}) is zero for $|y-z|$ much larger than
the lattice unit. We get, therefore
\begin{equation}
\label{29}
0 = \sum_n  2m_q \langle  P^a(n) A_\nu^b(y) V_\rho^c(z)\rangle  +
i f^{abd} \langle V_\nu^d(y)V_\rho^c(z)\rangle  +
i f^{acd} \langle A_\nu^b(y)A_\rho^d(z)\rangle \,, 
\end{equation}
where $P^a(n)$ is defined in eq.~(\ref{12}).

Consider finally the Ward identity eq.~(\ref{11}) with ${\cal O}(x)= P^b(x)$
as in Section~4 but do not sum over $n$. Assume that $x-n$ is much larger than
the lattice unit. Using similar steps as above we obtain
\begin{equation}
\label{30}
\langle P^b(x) {\bar \nabla}_\mu A_\mu^a(n)\rangle  = \langle P^b(x) 2 m_q P^a(n)\rangle \,.
\end{equation}
From  eq.~(\ref{30}) we conclude that ${\bar \nabla}_\mu A_\mu^a$ and
$2m_q P^a(n)$ have the same renormalization factor. Eq.~(\ref{29}) gives then
\begin{equation}
\label{31}
Z_A = Z_V = 1 \,.
\end{equation} 
We remark that in the considerations above the $m_q \rightarrow 0$ limit was
not necessary.

\section{Mixing}
In the study of weak, non-leptonic matrixelements it is essentially important
to construct operators in definite chiral representations. In the case of
Wilson fermions the nominal (tree level) chiral assignment of operators is
invalidated by quantum corrections due to the chiral symmetry breaking terms
in the action. Operators with definite chiral properties are linear
combinations of the operators with nominal assignment (mixing). For present
Monte Carlo calculations, as the case of the $B_K$ parameter shows,
the mixing coefficients should be calculated non-perturbatively, which is a 
highly non-trivial problem.

Using the example of local k-fermion operators (k=4,6,\dots) we are going to
demonstrate now that k-fermion operators in a definite chiral representation
(assignment on the tree level) do not mix with other k-fermion operators from
a different chiral representation. In particular, for the $\triangle s=2$,
4-fermion operator, whose matrixelement between ${\rm K}^0$ and 
${\bar {\rm K}}^0$ defines the $B_K$ parameter, the nominal chiral assignment
is equal to the full quantum chiral assignment -- there is no mixing.

Consider the Ward identity eq.~(\ref{11}) with 
${\cal O} \rightarrow {\cal O}(x) B(y_1,\dots,y_j)$, where ${\cal O}(x)$ is
some k-fermion operator in a definite chiral representation,
$B(y_1,\dots,y_j)$ is some product of local operators, and
$|y_1-x|,\dots,|y_j-x|$ are much larger than the lattice unit. 
Choosing the point
$n$ in  eq.~(\ref{11}) equal to $x$, or, if the operator ${\cal O}$ has an
extension of $O(a)$, summing over $n$ in the neighbourhood of $x$, then
on the l.h.s. the chiral
variation of the k-fermion operator ${\cal O}(x)$ enters. This is a linear
combination of k-fermion operators in the same representation. On the r.h.s.,
the $\langle X^a(n){\cal O}(x) B(y_1,\dots,y_j) \rangle$ matrixelement might
also produce k-fermion terms local in $x-n$ in different representations. This
is the source of mixing when using Wilson fermions \cite{CCH}. In our case
\begin{eqnarray}
\label{32}
\lefteqn {\langle X^a(n){\cal O}(x) B(y_1,\dots,y_j) \rangle =} \\
& &\langle \left(
\bar{\psi}_n \tau^a ({\hat h}\gamma^5 R {\hat h} \psi )_n + 
( \bar{\psi}{\hat h}\gamma^5 R {\hat h} )_n \tau^a \psi_n \right)
{\cal O}(x) B(y_1,\dots,y_j) \rangle \,. \nonumber 
\end{eqnarray}
If one of the fermions in $X^a$ are paired with one of the fermions in $B$,
the propagator is cancelled by $\hat {h}$ in $X^a$ and the result will be zero
since $n \sim x$ is far (in lattice units) from $y_1,\dots,y_j$ and $R$ 
and ${\hat h }$ are
local. Hence, both fermions in $X^a(n)$ should be paired with the fermions in
the k-fermion operator ${\cal O}(x)$. The result is a (k-2)-fermion operator
local in the point $x$. The dangerous term, therefore can not give a
k-fermion contribution to the l.h.s. of eq.~(\ref{11}). In the example of the
$B_K$ parameter, the matrixelement in eq.~(\ref{32}) is zero, since no
2-fermion operator exists with $\triangle s=2$.


\section{Acknowledgments}
The author is indebted for discussions with T.~DeGrand, J.~I.~Latorre,
H.~Leutwyler, H.~Neuberger
and F.~Niedermayer.

 
\newcommand{\PL}[3]{{Phys. Lett.} {\bf #1} {(19#2)} #3}
\newcommand{\PR}[3]{{Phys. Rev.} {\bf #1} {(19#2)}  #3}
\newcommand{\NP}[3]{{Nucl. Phys.} {\bf #1} {(19#2)} #3}
\newcommand{\PRL}[3]{{Phys. Rev. Lett.} {\bf #1} {(19#2)} #3}
\newcommand{\PREPC}[3]{{Phys. Rep.} {\bf #1} {(19#2)}  #3}
\newcommand{\ZPHYS}[3]{{Z. Phys.} {\bf #1} {(19#2)} #3}
\newcommand{\ANN}[3]{{Ann. Phys. (N.Y.)} {\bf #1} {(19#2)} #3}
\newcommand{\HELV}[3]{{Helv. Phys. Acta} {\bf #1} {(19#2)} #3}
\newcommand{\NC}[3]{{Nuovo Cim.} {\bf #1} {(19#2)} #3}
\newcommand{\CMP}[3]{{Comm. Math. Phys.} {\bf #1} {(19#2)} #3}
\newcommand{\REVMP}[3]{{Rev. Mod. Phys.} {\bf #1} {(19#2)} #3}
\newcommand{\ADD}[3]{{\hspace{.1truecm}}{\bf #1} {(19#2)} #3}
\newcommand{\PA}[3] {{Physica} {\bf #1} {(19#2)} #3}
\newcommand{\JE}[3] {{JETP} {\bf #1} {(19#2)} #3}
\newcommand{\FS}[3] {{Nucl. Phys.} {\bf #1}{[FS#2]} {(19#2)} #3}

\eject


\begin{thebibliography}{99}
\bibitem{4}
Some of the results presented here has been discussed in the first ref. of
\cite{PH}.
\end{thebibliography}
\end{document}